\documentstyle[prl,aps,epsfig,preprint]{revtex}
\begin{document}
\tighten

\title{A Modern Anatomy of Electron Mass}

\author{Xiangdong Ji and Wei Lu}
\bigskip

\address{
Department of Physics \\
University of Maryland \\
College Park, Maryland 20742 \\
USA\\
{~}}

\date{UMD PP\#98-093 ~~~DOE/ER/40762-145~~~ February 1998}

\maketitle

\begin{abstract}
Motivated by the need to understand hadron masses,
we reexamine an old problem in QED---the composition of the 
electron mass---in a modern perspective. We find that in the 
unrenormalized  QED,  the vacuum subtraction plays an important
role in understanding various  sources of the electron mass. 
The same issue  is also discussed in  the modified 
minimal subtraction scheme  with an emphasis on the  
scale and scheme dependence in the analysis. 
\end{abstract}
\pacs{xxxxxx}

\narrowtext

Understanding the internal structure of hadrons, the nucleon
in particular, is perhaps one of the most challenging problems
in theoretical physics. Part of the difficulty lies in
that hadrons are elementary excitations of the quantum
chromodynamics (QCD)
vacuum and hence involve infinitely many degrees of freedom 
from  the start.   Little is known about such systems. 
In particular, it is not  at all clear  that one can 
model the type of systems with just a few quantum 
mechanical degrees of freedom,  although such practices 
have been successful phenomenologically.
 
The topic of this paper is  the electron mass,
a subject that  was  once  the center of debate 
when quantum electrodynamics (QED) was being 
established. Over the years,  however, its importance has been 
superseded by the notion of mass renormalization,
which essentially means that the electron mass
is a fundamental observable requiring no further 
fundamental explanation. 
At first thought, the electron mass has little to do 
with the hadron structure problem stated above. 
As we shall demonstrate below, however, one 
can actually learn about some aspects of the field
theoretical bound states by reexamining 
this old issue.

From the  formal side of quantum field 
theory, one can see a number of similarities between 
the nucleon and electron. First of all, both of them 
are elementary excitations of the field
theoretical vacua, which show up as the poles of
the relevant Green's functions. Second, 
both are made of degrees of freedom which 
are the building blocks of the Lagrangian and 
physical observables but are not themselves 
observable. In the case of the nucleon, 
they are the bare or renormalized quarks and 
gluons, while in the case of the electron, 
they are the bare or renormalized 
electrons and photons which carry
the same quantum numbers as physical particles.
Finally, both the electron and nucleon are stable, and
have no natural description as bound 
states of known physical particles, unlike 
the positronium or deuteron. 

Needless to say, 
the detailed dynamics governing the nucleon and 
electron structure are very different. 
The nucleon is neutral in gauge charges, 
and is a strongly coupled
infinite-body system, whereas the electron is 
a charged particle whose structure
can be investigated in perturbation theory. 
Hence, we hope to learn 
some field theoretical aspects of the nucleon---those 
independent of dynamic details---through studying the 
simpler electron.

What we would like to emphasize here is  
a basic point of field theory, i.e., it is the same
set of degrees of freedom that  determines  
the vacuum and bound state structure.
To understand physical observables
of a bound state, one must know both the bound 
state and vacuum wave functions. Therefore, there exists 
{\it a priori} no set of degrees of freedom 
that controls the structure of bound states but  
has no effects on the vacuum. For an annihilation 
operator $b$, we have in general
\begin{equation}
     b|0\rangle \ne 0 \ , 
\end{equation}
except in free field theories. When an 
observable has an expectation value 
in the vacuum, one must always subtract this 
contribution in making physical 
predictions. Thus certain operators that
appear to be  positive-definite in Hilbert space may
have negative physical expectations.
For this reason, field theoretical results
may not always be conveniently mocked up by 
a system with a few quantum mechanical 
degrees of freedom. At this point, let us just 
mention that in both fields of hadron structure
and heavy-ion collisions, we have found in 
many papers  that  Eq. (1) 
is violated by  the ``quark and gluon" degrees of freedom. 

The subject of the electron mass 
has been an interesting problem historically.
In classical electrodynamics, the mass
of an electrically-charged particle presents
a mystery. When the particle is at rest 
and without spin, its total mass will include its 
electrostatic energy. As is well known, however, it 
diverges linearly as the radius of the particle 
goes  to zero. On the other hand, the physical 
mass of the object is apparently finite.

A consistent formulation of the electron mass
problem   became possible only after Dirac proposed 
his positron theory \cite{Dirac} -- the precursor of QED. 
According to its modern interpretation, the basic building 
blocks in the theory are the bare electrons
and photons, which are defined  when  the electromagnetic
interactions  {\it were}   turned off. 
Clearly then, they are not physically observable 
particles. The bare electron does have a  mass 
(the bare mass),  but it has  no electromagnetic origin. When the 
eletromagnetic interaction is turned on, the 
QED vacuum becomes nontrivial and the vacuum 
excitations produce a physical particle whose 
quantum numbers are the same as those of the bare 
electron. Pictorially this physical electron contains 
a bare electron in the QED vacuum 
plus the vacuum polarizations. 
Because of the small electromagnetic coupling, 
the structure of the physical electron can be calculated perturbatively 
using the bare degrees of freedom. Hence the physical 
mass of the electron can be ``explained'' in terms of 
the bare electron mass plus the contributions from the 
electromagnetic interactions.  In light of the fact 
that  the physical mass is observable while the bare mass 
is not, such an  explanation is not practically interesting 
and is usually ignored in modern textbooks.  
However, as we shall advocate in this paper, 
the explanation may help to understand some 
interesting aspects of the nucleon bound state
about which we know very little  in QCD.

According to the standard textbook formulation of QED, 
the physical electron mass is defined as 
the pole position of the full electron propagator \cite{Bj}. 
To order ${\cal O} ( e^2)$, the physical mass 
($m_P$)  is the  bare mass ($m$) 
plus a contribution from the self-energy diagram: 
\begin{equation} 
m_P=m +  \frac{1}
{2m} \times 
e^2  \int \frac{d^4 k}{(2\pi)^4} \bar u(P) 
\frac{\gamma_\alpha (\rlap/k + m)\gamma^\alpha}
{ (k^2 -m^2 + i \epsilon ) [(p-k)^2  + i \epsilon] }  u(P)  \ . 
\end{equation}  
Throughout this paper,  the electron mass ($m$), the coupling 
($e$) and  the operators  without further specification  
refer to the bare quantities.  To understand the  physical content
of this self-energy contribution,  one may  integrate out 
$k^0$ by contour method, 
\begin{equation}
m_P=m + 
\frac{e^2}{2} \int \frac{d^3{\bf k}}  {(2\pi)^3}
 \frac{m}{|{\bf k}|^2  E_{\bf k}  } 
-
\frac{e^2}{2m} \int \frac{d^3{\bf k} }  {(2\pi)^3}\Big(
 \frac{1}{E_{\bf k}  }- \frac{1}{|{\bf k}|}\Big) \ , 
\end{equation} 
where  $E_{\bf k}= \sqrt{ |{\bf k}|^2 +m^2|} $.
This result was  first obtained in the positron theory
by Weisskopf in  the early 1930's \cite{Weisskopf}. 
Quite importantly,  he observed that the electromagnetic energy 
is logarithmically divergent to all orders in 
perturbation theory, in contrast to 
the linear divergence in the mass of a classical 
charged particle.  Weisskopf also gave a detailed
explanation of this self-energy in the spirit of 
the linear response theory, in which it corresponds to 
the adiabatic switching on of the 
electromagnetic interactions.
He further classified the energy 
into the Coulomb energy (logarithmically divergent), 
the spin contribution (quadratically divergent), 
and the vacuum fluctuation energy (quadratically divergent). 
While such a classification is interesting 
in perturbation theory, we will attempt to 
understand the electron mass in a nonperturbative
formalism extendible to the mass of the nucleon. 
                
To this end, we consider the Hamiltonian formulation 
of QED. The physical mass is calculated as the 
expectation value of the Hamiltonian in the physical electron
at rest:
\begin{equation}
m_P=\frac{\langle P|H|P\rangle} 
{\langle P|P\rangle} \Big|_{\rm rest~ frame} \equiv
\langle H \rangle \ ,   
\end{equation} 
where we introduce the shorthand $\langle ~ \rangle$ for 
later convenience and the vacuum subtraction is
implied. The   QED Hamiltonian, 
in terms of the bare quantities, is well known: 
\begin{equation} 
H= H_e  + H_\gamma  + H_m
\end{equation} 
with  
\begin{eqnarray}
H_e&=& \int d^3 {\bf x}
~ \psi^\dagger (-i{\bf D} \cdot \alpha)\psi\ ,  \\
H_\gamma&=& \int d^3 {\bf x}
~\frac{1}{2} ({\bf E}^2 +{\bf B}^2) \ , \\
H_m&=& \int d^3 {\bf x} ~ m \bar \psi \psi  \  ,
\end{eqnarray}
where $\alpha$ is the 
Dirac matrix;  ${\bf E}$ and ${\bf B}$ are the 
electric and magnetic field strengths, respectively.  
The above decomposition of the QED Hamiltonian  
indicates that the electron energy 
consists of the kinetic energy (canonical plus 
current interaction energy), the 
electromagnetic energy, and the contribution 
from the  electron scalar density.  Notice that  each of these 
contributions is  independent of  gauge choices.

Accordingly,  the  physical electron mass may be 
written  as a sum  of three contributions,
\begin{equation}
      m_P = \langle H_e\rangle + \langle H_\gamma \rangle
        + \langle H_m \rangle \ . 
\end{equation} 
Evaluating  these matrix elements in perturbation
theory  to ${\cal O} (e^2)$, we obtain: 
\begin{eqnarray} 
\langle  H_e \rangle &=&
-\frac{e^2}{m} \int \frac{ d^3{\bf k} }  {(2\pi)^3} \Big(
 \frac{1}{E_{\bf k}  }- \frac{1}{|{\bf k}|}\Big)
-e^2 \int \frac{d^3{\bf k}}  {(2\pi)^3} 
\frac{m}{ E_{\bf k}^3}  \ , \label{10} \\
\langle  H_\gamma \rangle 
 &=&
\frac{e^2}{2} \int \frac{d^3{\bf k}}  {(2\pi)^3}
 \frac{m}{|{\bf k}|^2 E_{\bf k}  }  \ ,  \label{11} \\    
\langle  H_m \rangle &=&
m + \frac{e^2}{2m} \int \frac{ d^3{\bf k} }  {(2\pi)^3} \Big(
 \frac{1}{E_{\bf k}  }- \frac{1}{|{\bf k}|}\Big)
+e^2 \int \frac{d^3{\bf k}}  {(2\pi)^3}
\frac{m}{ E_{\bf k}^3}  \ . \label{17} 
\end{eqnarray}
Obviously,  the sum of  the above three matrix elements 
yields the same expression obtained from  the 
one-loop  self-energy diagram, Eq. (3).

Now we set about examining  the physical significance
of  the  individual contribution to the electron mass.
$ \langle  H_e  \rangle $ includes the canonical 
kinetic  and  current interaction energies 
of the internal bare electron.  We group them together
because separately they are not gauge invariant. In the
Feynman gauge,   we find  that the canonical kinetic energy is 
\begin{equation}
    \left\langle \int  d^3 {\bf x} 
~ \psi^\dagger \left(-i\vec{\alpha} 
      \cdot \vec{\partial}\right)\psi \right\rangle = 
\frac{e^2}{2m} \int \frac{ d^3{\bf k} }  {(2\pi)^3} \Big(
 \frac{1}{E_{\bf k}  }- \frac{1}{|{\bf k}|}\Big)
-e^2 \int \frac{d^3{\bf k}}  {(2\pi)^3}
\frac{m}{ E_{\bf k}^3} \ . 
\end{equation}
Using a momentum cutoff $\Lambda$, the above  expression 
can be simplified  to $-(10\alpha/4\pi)\log(\Lambda/m)$
in the leading logarithmic approximation, where
and henceforth $\alpha=e^2/4\pi$.  
The negative kinetic energy is  a consequence 
of the vacuum subtraction rather than the gauge 
choice. The  minus sign reflects the Pauli exclusion
principle which forbids some excitations of the
vacuum in  the presence of the bare electron.  In a model 
without the vacuum, the negative kinetic energy would be difficult 
to understand. In the same gauge, 
the current interaction energy is 
positive, but the  total gauge-invariant 
contribution $\langle H_e\rangle$ is still negative, 
reading  $-(\alpha/\pi) \log (\Lambda/m)$ to logarithmic accuracy. 
 
The electromagnetic energy, $ \langle  H_\gamma   \rangle $, 
can be further separated 
in a gauge invariant manner into the Coulomb 
and radiation contributions.  At ${\cal O}  (e^2)$  it is easy to show 
\begin{eqnarray}
\frac{1}{2} \langle  {\bf E}^2_{\rm Coul} \rangle &=& 
\frac{e^2}{2} \int \frac{d^3{\bf k}}  {(2\pi)^3}
 \frac{m}{|{\bf k}|^2 E_{\bf k}  } \ ,    \\
\frac{1}{2}\langle  {\bf E}^2_{\rm rad} +  
    {\bf B}^2  \rangle &=& 0 \ . 
\label{20}
\end{eqnarray}
Therefore,  the entire  electromagnetic energy 
comes from the Coulomb energy, whereas
those from the electric and magnetic
radiation fields cancel completely. While the Coulomb
energy is what one expects in a classical
picture, the cancellation of the radiative
contributions is somewhat surprising.  
We do not know whether this  persists 
to higher orders. Again, we would like
to point out that the radiation energy is 
formally positive-definite, but positivity 
is not guaranteed because of the vacuum subtraction.

Interestingly, the sum of the two contributions
in Eqs. (\ref{10}) and (\ref{11}) is free of ultraviolet divergences
${\cal O} (e^2)$. A simple calculation shows
\begin{equation}
    \langle H_e+H_\gamma\rangle = {3\alpha \over 2\pi} m\ .  
\end{equation}
This result is consistent with the fact
that $m\bar \psi\psi$ is a finite composite operator
at leading order.

The last component in Eq. (9), 
$ \langle  H_m   \rangle $, which is logarithmically divergent, 
is a product of the bare electron mass and scalar density.
The usual mass renormalization is to assume
that the bare mass cancels this logarithmic divergence 
so that the physical mass is finite.
The enhancement of the scalar density in the presence of 
the electromagnetic interactions is  essentially 
a relativistic effect. For comparison,  
we note that in nonrelativistic theory the scalar
density coincides with the charge density
which needs no renormalization
because of the current conservation.

Since bare degrees of freedom are not 
physical observables, one is not forced to
use them to formulate a physical calculation. 
Indeed, the very concept of renormalization was invented
on the basis of  this observation. In a
renormalizable theory, physical observables are 
calculated in terms of renormalized parameters, 
still unphysical though they are,  so that  no explicit 
divergences are present. In the remainder of this
paper, we  revisit  the electron mass 
from the point view of the  renormalized QED.
One must keep in mind, however, that
for  a renormalizable theory there exists an 
infinite number of renormalization 
schemes which are equivalent to each other.  
For QED, the most natural scheme  seems to be 
the on-shell one in which the renormalized
fields assume the  physical masses.  
For studying the mass structure of the electron, 
however, the usual prescription of the 
on-shell scheme is insufficient because one needs
additional renormalization conditions to define 
the composite operators with  non-vanishing anomalous dimensions.
Lacking in a natural extension of the on-shell scheme, 
we choose to work with  the 
modified minimal subtraction (${\overline {\rm MS}}$) scheme.
To regularize the divergences, we employ the dimensional 
regularization.

The starting point now is the renormalized version
of the QED Hamiltonian which is expressed in terms
of renormalized degrees of freedom (we will call them
the ${\overline {\rm MS}}$ electron and photon). 
Because the Hamiltonian is related to  the (00) component 
of the energy-momentum tensor $T^{\mu\nu}$, it is more 
convenient   to  consider  first the renormalization 
properties of the energy-momentum tensor. 
For a generic  field theory, one can  write down  its 
energy-momentum tensor as the Noether current corresponding  
to space-time translations.  By adding appropriate 
surface terms, one can obtain a symmetric tensor after 
using the equations of motion.  Such a procedure is called 
the Belinfante  improvement \cite{Belin}. 
The symmetric form of the energy-momentum 
tensor can be split into  a sum of the traceless and trace parts: 
\begin{equation}
T^{\mu\nu}= \bar T^{\mu\nu} +\hat T^{\mu\nu} \ ,    
\end{equation}
where 
\begin{eqnarray}
\bar T^{\mu\nu}&=& T^{\mu\nu} -\frac{1}{4} g^{\mu\nu} T^\alpha_{~\alpha} \ ,
 \\   
 \hat  T^{\mu\nu}&=& \frac{1}{4} g^{\mu\nu} T^\alpha_{~\alpha} \ . 
\end{eqnarray} 
Because $T^{\mu\nu}$ is a symmetry current, it is 
a finite operator requiring no renormalization.  Furthermore,
different representations of the Lorentz group 
do  not mix under renormalization, so 
both  $\bar T^{\mu\nu}$ and $\hat T^{\mu\nu}$ are independent
of renormalization scale and scheme.

The above separation of $T^{\mu\nu}$, combined with  
\begin{equation}
    \langle P | T^{\mu\nu}|P \rangle = 2P^\mu P^\nu \ , 
\end{equation} 
implies a {\it Virial} theorem  \cite{Ji} for the mass of
a particle: 3/4 of the physical mass 
comes from the traceless part of the 
energy-momentum tensor while the remaining  1/4 comes  from the trace part. 
Notice that such a  partition  of the mass is valid for 
any  $physical$ particle---composite or point-like.  

Now we consider the QED energy-momentum 
tensor operator in detail. We  work in the covariant gauge.  
The gauge-fixed Lagrangian reads  
\begin{equation} 
{\cal L}= \bar \psi (i \stackrel{\leftrightarrow}{\not\! D} -m) \psi 
-\frac{1}{4} F^{\alpha\beta} F_{\alpha\beta} 
-\frac{1}{2 \xi} (\partial \cdot A)^2   \ ,
\end{equation} 
where $\stackrel{\leftrightarrow}{D^\mu}=
\frac{1}{2}(\stackrel{\leftarrow}{D^\mu}+ 
\stackrel{\rightarrow}{D^\mu})$ with $  \stackrel{\rightarrow}{D^\mu}
=\stackrel{\rightarrow}{ \partial^\mu}  +i e A^\mu$
and $\stackrel{\leftarrow}{D^\mu} = 
- \stackrel{\leftarrow}{\partial^\mu} +i e A^\mu$, 
and $\xi$ is the gauge fixing parameter.  
The  Belinfante-improved  energy-momentum tensor 
can be easily derived:
\begin{eqnarray} 
T^{\mu\nu}& = &-g^{\mu\nu}{\cal L} 
 -F^{(\mu\alpha}F^{\nu)}_{~\alpha}
+\bar \psi i \stackrel{\leftrightarrow}{D^{(\mu}} 
 \gamma^{\nu)} \psi
\nonumber \\
& &
-g^{\mu\nu}\xi^{-1} \partial^\alpha (A_\alpha \partial \cdot A)
+2 \xi^{-1} A^{(\mu} \partial^{\nu)} (\partial \cdot A) 
\nonumber \\ 
& & 
-\frac{\delta S}{\delta A_\mu} A^\nu 
-\frac{1}{8}\frac{\delta S}{\delta \psi}[\gamma^\mu, \gamma^\nu] \psi 
-\frac{1}{8}\bar\psi [\gamma^\mu, \gamma^\nu] \frac{\delta S}{\delta \bar \psi}
 \ ,  \label{pond}
\end{eqnarray} 
where $(\mu\nu)$ stands for symmetrization of the indices  and 
$S\equiv \int d^4  x {\cal L}$ is the action.
It is straightforward to show that 
\begin{equation}
\partial_\mu T^{\mu\nu} = 
-\frac{\delta S}{\delta A_\mu} \partial^\nu A^\mu 
-\frac{\delta S}{\delta \psi} \partial^\nu \psi 
 -\frac{\delta S}{\delta \bar \psi} \partial^\nu  \bar \psi \ .  
\end{equation} 
Thus   $T^{\mu\nu}$  is  symmetric and conserved  
after using the  equations of motion. 
In the following discussion, 
we  suppress  the gauge-fixing and 
equation-of-motion related operators 
because they do not contribute  
in the physical matrix elements \cite{Lee}.

For the traceless part, we  write
\begin{equation}
\bar T^{(\mu\nu)} = \bar T^{(\mu\nu)}_1 +\bar T^{(\mu\nu)}_2   \ ,  
\end{equation} 
where 
\begin{eqnarray}
\bar T^{(\mu\nu)}_1 &=& \bar \psi i
\stackrel{\leftrightarrow}{D^{(\mu}}
\gamma^{\nu)} \psi 
 \ ,  \\
\bar T^{(\mu\nu)}_2 &=& -  F^{(\mu\alpha} F^{\nu)}_{~\alpha}
\ , 
\end{eqnarray}
are the electron covariant kinetic  energy density
and electromagnetic  energy density, respectively. 
Under renormalization, $T^{\mu\nu}_1$ and $T^{\mu\nu}_2$ 
mix with each other, but their  sum is invariant \cite{Ji}:
\begin{equation}
\bar T^{(\mu\nu)}= \bar T^{(\mu\nu)}_{1,R} (\mu) 
+ \bar T^{(\mu\nu)}_{2,R} (\mu) \ .    
\end{equation} 
According to the Lorentz symmetry,   the matrix elements
of  these two operators can be parameterized  as follows:  
\begin{eqnarray} 
\langle  P|  \bar T^{(\mu\nu)}_{1,R}  (\mu)|P \rangle 
&=&2 a(\mu)   \Big(P^\mu P^\nu -\frac{1}{4}g^{\mu\nu} m^2_P \Big) \ , \\ 
\langle  P| \bar T^{(\mu\nu)}_{2,R}  (\mu)|P \rangle
 &= &2\left(1- a(\mu)\right)
  \Big(P^\mu P^\nu -\frac{1}{4}g^{\mu\nu} m^2_P \Big) \ ,
 \end{eqnarray} 
where $a(\mu)$ is a scale-and-scheme 
dependent parameter. 

At the classical level,  the nonvanishing electron  mass 
implies  the  breaking of scale symmetry  in the QED  
Lagrangian. After quantization, the 
scale symmetry is further broken by radiative 
effects. As a consequence, an anomaly 
arises in the trace part of the energy-momentum 
tensor in the course of  renormalization.
According to Ref. \cite{Collins},  
the trace part of the energy-momentum tensor can be 
effectively taken  as a sum of two  renormalized   operators:  
\begin{equation} 
T^\alpha_{~ \alpha }= O_{m,R}(\mu)  + O_{a,R} (\mu) \ , 
\end{equation} 
where 
\begin{eqnarray} 
O_{m,R} &=&(1 + \gamma_m) m_R (\bar \psi \psi)_R \ ,  \\ 
O_{a,R} &=& \frac{\beta}{2 e_R} \left(F^{\alpha\beta} 
F_{\alpha\beta}\right)_R \ .  
\end{eqnarray} 
In the above expression, 
  $\beta/e_R =\alpha_R/(3 \pi) + \cdots $  is the QED 
beta function  and  $\gamma_m= 3\alpha_R/ (2\pi) + \cdots $  
the anomalous dimension of the electron mass term.
The occurrence  of $O_{a, R}$ is  a reflection 
of the scale symmetry  breaking and the term is called  
trace anomaly. The matrix elements of $O_{m, R}$  and $O_{a,R}$ 
can be parameterized respectively as 
\begin{eqnarray} 
\langle P| O_{m,R}  (\mu)| P\rangle &=&  2 b(\mu) m^2_P \ , 
 \\ 
 \langle P| O_{a,R}  (\mu)| P\rangle &=&  2 [1- b(\mu)] m^2_P \ ,  
\end{eqnarray} 
 where $b(\mu)$ is  a scale-and-scheme dependent parameter.

According to the foregoing  discussion,  we obtain  
the following renormalized QED Hamiltonian
\begin{eqnarray} 
H&\equiv & \int d^3 {\bf x} T^{00} (0, {\bf x}) \nonumber  \\
 &=& H_{e,R} (\mu) + H_{\gamma,R} (\mu)  +H_{m,R} (\mu)  
+ H_{a,R} (\mu)\ ,  \label{36} 
\end{eqnarray} 
where $H_{e,R} $ and $H_{\gamma,R} $ are  the renormalized  versions 
of  $H_{e}$ and $H_{\gamma}$,  respectively, and 
\begin{eqnarray} 
H_{m,R}&=&\int d^3 {\bf x}
\Big( 1 +\frac{\gamma_m}{4}\Big)
m_R (\bar \psi  \psi)_R  \ , \\  
H_{a,R} &=&\int d^3 {\bf x} \frac{\beta}{4 e_R}
({\bf E}^2 -{\bf B}^2 )_R \ . 
\end{eqnarray}
Applying  Eq. (\ref{36}) to the physical  electron  state 
at rest,  one can attain the  following 
partition of the  physical electron mass 
among various sources: 
\begin{equation} 
m_P= m_e (\mu)  + m_\gamma  (\mu)  + m_m (\mu) 
 + m_a  (\mu)  \ . 
\end{equation} 
In relation to the matrix elements of the energy-momentum tensor, 
there are
\begin{eqnarray} 
m_e  (\mu)  &=& \frac{3}{4} \left[ a(\mu) -\frac{b(\mu)}
{1 +\gamma_m(\mu) }\right] m_P \ ,  \label{40} 
 \\ 
m_\gamma  (\mu) &=& \frac{3}{4} \left[1-a(\mu) \right]m_P \ ,  
 \\ 
m_m  (\mu) &=&\frac{b (\mu) }{4} 
\left[1 + \frac{3}{1 +\gamma_m(\mu)}\right] 
 m_P \ ,  
 \\ 
m_a  (\mu) &=& \frac{1}{4} \left[1-b (\mu) \right] m_P \ . \label{43} 
\end{eqnarray} 

At  the one-loop level, explicit calculations show that 
\begin{equation} 
a(\mu)=1 -\frac{2\alpha_R(\mu)} {3 \pi} 
\left[\log\frac{\mu}{m^2_R(\mu)} + \frac{17}{12}\right] \ .
\label{44}  
\end{equation} 
The  photon anomaly operator $ O_{a,R}$ begins  to   contribute 
at  ${\cal O} (e^4_R)$,  so   its effect can  simply  be neglected in our 
${\cal O} (e^2_R)$ analysis.  In other words,  we have 
\begin{equation} 
b(\mu)= 1 +{\cal O}( \alpha^2_R (\mu) ) \ . \label{45} 
\end{equation} 
Substituting Eqs. (\ref{44}) and (\ref{45}) into 
(\ref{40})--(\ref{43}),  we obtain 
\begin{eqnarray} 
m_e (\mu)  &=&- \frac{\alpha_R(\mu)} {2 \pi}
\left[\log\frac{\mu^2}{m^2_R(\mu)} - \frac{5}{6}\right]m_R (\mu) 
+  {\cal O}( \alpha^2_R (\mu^2) ) \ , \\ 
  m_\gamma (\mu)  &=& \frac{\alpha_R(\mu)} {2 \pi}
\left[\log\frac{\mu^2}{m^2_R(\mu)} + \frac{17}{12}\right]m_R (\mu)  
+  {\cal O}( \alpha^2_R (\mu) ) \ , \\ 
m_m  (\mu)    &=&\left[ 1 + \frac{3 \alpha_R (\mu) }{4 \pi}
\left(\log\frac{\mu^2}{m^2_R(\mu)} - \frac{1}{6}\right) \right]  m_R (\mu) 
+  {\cal O}( \alpha^2_R (\mu) ) \ ,  
 \\ 
 m_a  (\mu)   &=&  {\cal O}( \alpha^2_R (\mu) ) \ , 
\end{eqnarray} 
where $m_R(\mu)$ is the mass of  the electron field
in the $\overline {\rm MS}$ scheme. The logarithms in 
the above equations are consistent with our earlier results
in terms of  the  bare quantities. 

The physical electron is seen as made  of 
the $\overline{\rm MS}$ electrons and photons;  
the latter are,  of course,  unobservable.
Loosely speaking,  the renormalization scale $\mu$ defines
the sizes of these elementary degrees of freedom. To get a physically 
interesting picture of the electron, $\mu$ cannot 
be  much larger nor  smaller than the physical mass. 
If $\mu$ is too large, the $\overline {\rm MS}$
 electron is resolved  at an unnecessarily fine scale, 
and the $\overline {\rm MS}$ degrees of freedom become
similar to the bare ones. If $\mu$ is too small, 
much of the electron structure physics is 
embedded in the $\overline{\rm MS}$ degrees 
of freedom, and unwanted large infrared logarithms 
appear. An optimal choice for $\mu$ is 
around the physical mass of the electron. If setting $\mu=m_R(\mu)$, 
we have
\begin{eqnarray}
   m_e(\mu) &=& \left({5\alpha_R (\mu)  \over 12\pi}\right) m_R(\mu)  \ ,  \\
   m_\gamma(\mu)& = & \left({17\alpha_R (\mu) 
 \over 24\pi}\right) m_R(\mu)  \ , \\
   m_m(\mu) & = & \left(1-{3\alpha_R (\mu) 
\over 24\pi}\right)m_R(\mu) \ ,  \\
   m_a(\mu) & = & 0 . 
\end{eqnarray}
This decomposition gives some ``intuitive'' feeling
about how the physical electron is made of the ${\overline {\rm MS}}$
constituents. Equation (50) shows that the kinetic and 
current interaction  energy of the ${\overline {\rm MS}}$ electron 
is positive;  Eq. (52) indicates that the scalar density is now 
less than the charge density. A similar decomposition
of the nucleon mass  was  performed   in Ref. \cite{Ji}.

One cannot overemphasize the fact that any picture
of a bound state in quantum field theory is necessarily 
scheme- and scale-dependent. In this respect, the 
electron mass decomposition  in the  unrenormalized 
QED and  its parallel in the ${\overline {\rm MS}}$ 
scheme present   a sharp contrast.  While a choice 
of scheme and scale has to be made in performing 
practical  calculations, physical results are 
independent of such  choices. Nonetheless, 
a judicious choice of scheme may help us  
understand the physics  better than others and
motivate model building when an exact 
calculation is  not available. It is 
not clear, however, which choice of
scheme and scale in QCD, if any, may simplify 
the hadron structure problem to a quantum 
mechanical one with few degrees of freedom. 

This work is  supported by  the U.S. Department of Energy 
under grant number DE-FG02-93ER-40762.

\end{document}